\documentclass[twocolumn, preprintnumbers, superscriptaddress, showkeys]{revtex4-2}

\usepackage{amsfonts}
\usepackage{amssymb}
\usepackage{anysize}
\usepackage{appendix}
\usepackage{bm}
\usepackage{color}
\usepackage{dcolumn}
\usepackage{epsf}
\usepackage{epstopdf}
\usepackage{float}
\usepackage{graphicx}
\usepackage{mathrsfs}
\usepackage{mathtools}
\usepackage{multirow}
\usepackage{pstricks,pst-node,pst-text,pst-3d}
\usepackage{setspace}
\usepackage{subfigure}
\usepackage{booktabs}
\usepackage{caption}
\usepackage{subcaption}
\usepackage[all]{xy}

\DeclarePairedDelimiterX\braket[2]{\langle}{\rangle}{#1 \delimsize\vert #2}

\makeatletter
  \newcommand\figcaption{\def\@captype{figure}\caption}
  \newcommand\tabcaption{\def\@captype{table}\caption}
\makeatother

\renewcommand{\thetable}{\arabic{table}}

\begin{document}


\title{Reinforcement Learning for Charging Optimization of Inhomogeneous Dicke Quantum Batteries}
\author{Xiaobin Song}
\affiliation{CFINS, Department of Automation, Beijing National Research Center for Information Science and Technology, Tsinghua University, Beijing, 100084, China}

\author{Siyuan Bai}
 \thanks{Address correspondence to bbjiang@tsinghua.edu.cn, rbwu@tsinghua.edu.cn, and baisiyuan@lzu.edu.cn}
\affiliation{Key Laboratory of Quantum Theory and Applications of Ministry of Education, Lanzhou Center for Theoretical Physics, Gansu Provincial Research Center for Basic Disciplines of Quantum Physics, and Key Laboratory of Theoretical Physics of Gansu Province, Lanzhou University, Lanzhou 730000, China}

\author{Da-Wei Wang}
\affiliation{School of Integrated Circuits, Tsinghua University, Beijing, 100084, China}

\author{Hanxiao Tao}
\affiliation{CFINS, Department of Automation, Beijing National Research Center for Information Science and Technology, Tsinghua University, Beijing, 100084, China}

\author{Xizhe Wang}
\affiliation{CFINS, Department of Automation, Beijing National Research Center for Information Science and Technology, Tsinghua University, Beijing, 100084, China}

\author{Rebing Wu}
 \thanks{Address correspondence to bbjiang@tsinghua.edu.cn, rbwu@tsinghua.edu.cn, and baisiyuan@lzu.edu.cn}
\affiliation{CFINS, Department of Automation, Beijing National Research Center for Information Science and Technology, Tsinghua University, Beijing, 100084, China}

\author{Benben Jiang}
 \thanks{Address correspondence to bbjiang@tsinghua.edu.cn, rbwu@tsinghua.edu.cn, and baisiyuan@lzu.edu.cn}
\affiliation{CFINS, Department of Automation, Beijing National Research Center for Information Science and Technology, Tsinghua University, Beijing, 100084, China}

\begin{abstract}
Charging optimization is a key challenge to the implementation of quantum batteries, particularly under inhomogeneity and partial observability. This paper employs reinforcement learning to optimize piecewise-constant charging policies for an inhomogeneous Dicke battery. We systematically compare policies across four observability regimes, from full-state access to experimentally accessible observables (energies of individual two-level systems (TLSs), first-order averages, and second-order correlations). Simulation results demonstrate that full observability yields near-optimal ergotropy with low variability, while under partial observability, access to only single-TLS energies or energies plus first-order averages lags behind the fully observed baseline. However, augmenting partial observations with second-order correlations recovers most of the gap, reaching 94\%–98\% of the full-state baseline. The learned schedules are nonmyopic, trading temporary plateaus or declines for superior terminal outcomes. These findings highlight a practical route to effective fast-charging protocols under realistic information constraints.

\end{abstract}

\keywords{Inhomogeneous dicke quantum batteries; Charging optimization; Reinforcement learning; Partial observability}

\maketitle

\section{Introduction}
The increasing demand for efficient and rapid energy storage solutions at the nanoscale has spurred significant interest in the development of quantum technologies for energy management. Quantum batteries, devices that store energy in quantum mechanical degrees of freedom, have emerged as a promising paradigm, offering the potential for faster charging time and higher energy densities compared to their classical counterparts \cite{Alicki2013, Andolina2019a, Campaioli2024, Quach2022}. The key advantage of quantum batteries lies in their ability to exploit quantum phenomena, such as entanglement and coherence, to achieve a quantum advantage in charging power, a phenomenon known as supercharging \cite{Campaioli2017, Binder2015}.

The central challenge in harnessing the potential of quantum batteries lies in designing effective charging protocols \cite{Shang2025, Gemme2023}. The charging process is fundamentally a problem of quantum control: how to manipulate external parameters, such as driving fields, to steer the battery from a low-energy (discharged) state to a high-energy (charged) state in the shortest possible time while maximizing the stored energy and minimizing energy dissipation \cite{Andolina2019b}.

Traditionally, this problem has been tackled using the tools of optimal control theory \cite{Brif2010}. These methods seek to find a time-dependent control Hamiltonian that optimizes a specific objective function, such as the charging time or the final energy of the battery. Seminal works in this area have provided analytical solutions for simple quantum battery models, revealing fundamental limits on charging power and highlighting the role of quantum correlations in achieving charging speed-ups \cite{Rossini2020, Ferraro2018}. However, these approaches are often limited to idealized models with a small number of components and simple interactions. In addition to classical optimal control, recent work in the broader control community has explored learning-based approximations of optimization-based controllers, primarily to reduce online computational overhead \cite{Tagliabue2022, Pozzi2023, Pozzi2025}. In particular, imitation learning has been used to distill expert controllers such as stochastic model predictive control into neural surrogates that can be evaluated efficiently at deployment. Such approaches, however, rely on access to an expert optimizer and on generating representative expert trajectories. For quantum-battery charging, the dynamics are governed by high-dimensional Hamiltonian evolution, and obtaining expert solutions via solver-in-the-loop optimal control can become computationally prohibitive as the dimension grows.

To address these limitations, researchers have begun to explore the use of machine learning, and specifically reinforcement learning (RL) \cite{Sutton1998, Lillicrap2015, Haarnoja2018}, as a powerful alternative for discovering optimal charging protocols \cite{Erdman2024, Sun2024, Zakavati2025}. RL offers a model-free approach to control problems, where an agent learns an optimal policy by interacting with its environment and receiving feedback in the form of rewards. This data-driven approach is well-suited for the complexities of quantum systems, where it has been successfully applied to various tasks such as quantum state preparation and gate design \cite{Bukov2018a, Bukov2018b, Niu2019, Li2025}. It can potentially discover non-intuitive control strategies that are missed by conventional methods.

Recent work \cite{Erdman2024} applied reinforcement learning to optimize fast-charging protocols for Dicke quantum batteries with strong performance. However, the Dicke model considered there is homogeneous, and the agent has access to the full quantum state. These are idealizations that are rarely attainable in practice. Crucially, small imperfections in physical implementations would inevitably break the uniformity of the atomic frequencies as exemplified by Doppler shifts from atomic motion in cavity QED systems \cite{Zhiqiang:17, Dimer2007}. Consequently, real devices exhibit inhomogeneous parameters, and only limited, experimentally accessible observables are available for control.

Motivated by these gaps, this work explores reinforcement learning as a data-driven route to fast charging under such realistic conditions. We focus on an inhomogeneous Dicke quantum battery and cast charging as a discrete-time RL problem in which an agent learns a piecewise-constant control sequence for the charger–battery coupling within a bounded, continuous action space. The agent is guided by a dense, shaped reward that blends energy and ergotropy gains while the overall objective targets terminal ergotropy. To mirror experimental readout limits, we consider four observability regimes ranging from full-state access to accessible summaries built from single-TLS energies, first-order collective averages, and second-order correlation features. Policies are trained with Soft Actor–Critic \cite{Haarnoja2018, Haarnoja2019}, leveraging entropy regularization and replay-based off-policy updates. Together, these elements lay the groundwork for intelligent charging protocols that remain effective under partial observability and realistic readout constraints.

The rest of this work is organized as follows. In Section \ref{sec2} we introduce the Dicke quantum battery model. The reinforcement learning framework for charging optimization is stated in Section \ref{sec3}. The performance of the framework is evaluated and discussed in Section \ref{sec4}, followed by conclusions in Section \ref{sec5}.

\section{Dicke Quantum Battery  and Charging Dynamics }
\label{sec2}

A paradigmatic model for exploring collective effects in quantum batteries is the Dicke model\cite{Kirton2019}. In this framework, the quantum battery consists of $N$ two-level systems (TLSs), or qubits, which act as the battery cells. These cells are collectively coupled to a single bosonic mode, typically a cavity field, which serves as the charger. The collective nature of the interaction is the key ingredient that allows for the supercharging effect\cite{Ferraro2018}.

The Hamiltonian of the total system (battery plus charger) in the standard uniform Dicke model is given by \(  \hat{H} (t)=  \hat{H}_B +  \hat{H}_C +  \hat{H}_{\text{int}}(t) \). Here, \(  \hat{H}_B \) describes an ensemble of \( N \) identical battery cells, each modeled as a TLS with ground state \( |g\rangle \), excited state \( |e\rangle \), and an energy splitting \( \omega_0 \). Specifically, the battery Hamiltonian takes the form (in this paper we take $\hbar=1$)
\begin{equation}
    \hat{H}_B = \omega_0 (\hat{J}_z + J), \label{eq:Hb}
\end{equation}
where \( \hat{J}_z = \frac{1}{2} \sum_{j=1}^{N} \hat{\sigma}_z^{(j)} \) is the collective pseudo-spin operator, with \( \sigma_z^{(j)} \) denoting the Pauli-\( Z \) operator for the \( j \)-th cell. The total spin is represented by \( J = N/2 \), and the constant \( J \) in Eq. \eqref{eq:Hb} is introduced to ensure that the average energy remains positive, simplifying subsequent analysis. The charger subsystem is represented by a single-mode cavity with frequency $\omega_c$, and its Hamiltonian is given by
\begin{equation}
    \hat{H}_C = \omega_c \hat{a}^\dagger \hat{a},
\end{equation}
where $\hat{a}^\dagger$ and $\hat{a}$ denote the bosonic creation and annihilation operators of the cavity mode, respectively. Furthermore, we consider the resonant condition in which the cavity frequency matches the energy splitting of the TLSs, namely $\omega_c = \omega_0$. The energy exchange between the battery and the charger is governed by the interaction Hamiltonian $\hat{H}_{\text{int}}$. In the ideal Dicke model, all battery cells couple identically to the cavity mode, leading to the interaction Hamiltonian
\begin{equation}
    \hat{H}_{\text{int}}(t) = 2 \lambda_c(t) \hat{J}_x (\hat{a} + \hat{a}^\dagger),
\end{equation}
where $\hat{J}_x = \frac{1}{2} \sum_{j=1}^{N} \hat{\sigma}_x^{(j)}$ with $\hat{\sigma}_x^{(j)}$ being the  Pauli-\( X \) operator for the \( j \)-th cell, and $\lambda_c(t)$ denotes the uniform coupling strength, which serves as a classical external control parameter. It is noteworthy that we do not employ the rotating-wave approximation (RWA) here. This is because the RWA is typically valid only under weak coupling conditions, which may not hold for the arbitrary values of the control parameter $\lambda_c(t)$ considered in our study. 

The dynamics of the ideal Dicke model is confined to an $(N+1)$-dimensional subspace due to the conservation of the total spin operator $\hat{\mathbf{J}}^2$, a consequence of the system's global SU(2) symmetry, which is widely used in the study of Dicke quantum batteries~\cite{Erdman2024}. However, the assumption of identical cells represents an idealization. In realistic physical implementations, imperfections and manufacturing variations lead to non-uniformity in the properties of the individual TLSs. To capture this effect, we extend the Dicke model by allowing each cell to have a distinct energy splitting $\omega_j$. This non-uniformity breaks the collective symmetry of the standard Dicke model. 
The modified battery and interaction Hamiltonians become:
\begin{subequations} \label{eq:Hprime}
\begin{align}
\hat{H}'_B
    &= \sum_{j=1}^{N}
       \frac{\omega_j}{2}
       \left(
         \hat{\sigma}_z^{(j)} + 1
       \right), \label{eq:HprimeB}\\
\hat{H}'_{\mathrm{int}}(t)
    &= \lambda_c(t)
       \sum_{j=1}^{N}
       \hat{\sigma}_x^{(j)}
       \left(
         \hat{a} + \hat{a}^\dagger
       \right). \label{eq:HprimeInt}
\end{align}
\end{subequations}
Here, $\omega_j = \omega_0 + \delta_j$ denotes the transition frequency of the $j$-th TLS, where $\delta_j$ represents the disorder strength or deviation from the nominal frequency $\omega_0$.  The total Hamiltonian of the non-uniform system is given by $\hat{H}'(t) = \hat{H}'_B + \hat{H}_C + \hat{H}'_{\text{int}}(t)$.
The presence of disorder in both on-site energies and couplings renders the system's dynamics significantly more complex and generally intractable to analytical optimal control techniques. This complexity motivates the adoption of adaptive, model-free approaches such as reinforcement learning.

The charging process is initiated with the system in a well-defined initial state. The battery is assumed to be fully discharged, meaning all TLS are in their ground state. For the non-uniform model, this corresponds to the product state $|\psi(0)\rangle_{B} = |g_1, g_2, \dots, g_N\rangle$. The charger is prepared in a state with a large number of excitations, such as a Fock state $|n\rangle$ with $n = N$. The system then evolves according to the time-dependent Schrödinger equation:
\begin{equation}
    i \frac{d}{dt} |\psi(t)\rangle = \hat{H}'(t) |\psi(t)\rangle, 
\end{equation}
where  the initial state is $|\psi(0)\rangle = |\psi(0)\rangle_{B} \otimes |n\rangle$.

The key figure of merit is the instantaneous energy $E(t) = \sum_j E_j(t)$ stored in the battery at time $t$, where $E_j(t) = \langle \psi(t) |  \frac{\omega_j}{2} ( \hat{\sigma}_z^{(j)} +1 ) | \psi(t) \rangle$ is the instantaneous energy of the $j$-th TLS. Another important quantity characterizing the stored energy is the ergotropy $\mathcal{E}(t)$ at time $t$, defined as
\begin{equation}
    \mathcal{E}(t) = Tr[\hat{H}'_B \rho(t)] - Tr[\hat{H}'_B \tilde{\rho}(t)],
\end{equation}
where $\rho(t) \equiv Tr_C[|\psi(t)\rangle\langle\psi(t)|]$ is the reduced density matrix of the battery, and $\tilde{\rho}(t)$ is the corresponding passive state. This passive state is given by $\tilde{\rho}(t) = \sum_{i=1}^{N} \lambda_i |\epsilon_i\rangle\langle\epsilon_i|$, where $\{\lambda_i\}$ are the eigenvalues of $\rho(t)$ sorted in decreasing order ($\lambda_1 \ge \lambda_2 \ge \cdots \ge \lambda_N$), and $\{|\epsilon_i\rangle\}$ are the eigenstates of $\hat{H}'_B$ with energies sorted in increasing order ($\epsilon_1 \le \epsilon_2 \le \cdots \le \epsilon_N$). The ergotropy represents the maximum amount of energy that can be extracted from the quantum state via cyclic unitary operations\cite{Binder2015}. The objective is to find a control protocol for the coupling strength $\lambda_c(t)$ that maximizes the final ergotropy $\mathcal{E}(\tau)$ at a given time $\tau$.

\section{Reinforcement Learning Framework for Charging Optimization}
\label{sec3}

To determine the optimal time-dependent control function $\lambda_{c}(t)$, we reformulate the charging process as a discrete-time reinforcement learning problem. An RL agent interacts with a simulated quantum battery system as its environment to acquire a control policy. At each time step $k$, the agent observes a state $s_k$ and selects an action $a_k$ according to a deterministic or stochastic policy $\pi(a_k|s_k)$. The objective is to maximize the expected return, i.e., the cumulative reward over the charging horizon.

\subsection{Environment for the RL agent}
\label{sec:env}
The environment for the RL agent is the simulated or real inhomogeneous Dicke battery itself, composed of $N$ TLSs and a charger. It emulates the quantum dynamics induced by a given charging protocol. The total charging time $\tau$ is divided into $K$ discrete steps of duration $\Delta t = \tau / K$. At the beginning of each step $k(k=0,1,\ldots,K-1)$ , the agent selects an action $a_k$, which sets the coupling strength $\lambda_{c}$ to a constant value for the duration of that step. The environment then evolves the quantum state from $|\psi(t_k)\rangle$ to $|\psi(t_{k+1})\rangle$ by solving the Schrödinger equation with a piecewise-constant Hamiltonian:
\begin{equation}
    |\psi(t_{k+1})\rangle = \exp(-i \hat{H}'(a_k) \Delta t) |\psi(t_k)\rangle
\end{equation}
where $\hat{H}'(a_k)$ is the total system Hamiltonian with the coupling strength determined by $a_k$ .

The state $s_k$ should provide the agent with sufficient information about the system at time $t_k (t_k = k \Delta t) $ and satisfy the Markov property (i.e., $s_{k+1}$ must only depend on $s_k$ and the applied action $a_k$). While the full quantum state vector $|\psi(t_k)\rangle$ is information-complete, it is generally not accessible in experiments and thus impractical as an RL state representation. To address this limitation, we replace $s_k$ with experimentally accessible observables 
$o_k$ and evaluate whether such observations suffice to learn high-quality charging protocols. In this paper we consider three families of measurements: the expected energy of each TLS, first-order averages, and second-order correlations. Moreover, as in \cite{Erdman2024}, we also incorporate the last selected action $a_{k-1}$ and the current time step $t_k$ into the policy input for both the full-state and partial-observation settings.

Note that replacing the full state $s_k$ by a restricted observation $o_k$ does not guarantee that the resulting process is Markovian in the observation space. Experimentally accessible observables are typically not information-complete, so the mapping $o_k = f(|\psi(t_k)\rangle)$ is generally many-to-one. Consequently, distinct quantum states may correspond to the same $o_k$, and the transition to $o_{k+1}$ may also depend on additional, unobserved physical quantities beyond $o_k$ and $a_k$, rendering the control problem effectively a partially observable Markov decision process (POMDP) \cite{Kaelbling1998}. In this work we treat this setting with a practical baseline by learning a memoryless policy $\pi(a_k \mid o_k)$ augmented with $a_{k-1}$ and $t_k$, and we empirically assess the impact of partial observability by comparing a hierarchy of observation sets.

The action $a_k$ specifies the coupling strength $\lambda_c$ applied during step $k$. We define a bounded continuous action space:
\begin{equation}
    \mathcal{A} = [\lambda_{c,\text{min}},\lambda_{c,\text{max}}] \subset \mathbb{R}
\end{equation}
where the bounds $\lambda_{c,\text{min}}$ and $\lambda_{c,\text{max}}$ are hyperparameters informed by physical knowledge. At each step $k$ the agent selects $a_k \in \mathcal{A}$ and holds it fixed for the duration of that step (i.e., $[t_k,t_{k+1})$), thereby realizing a piecewise-constant charging protocol.

The reward function $R(s_k, a_k, s_{k+1})$ is crucial for guiding the agent towards the desired charging performance. To encourage efficient charging, we define the step-$k$ reward as the increment in the battery’s ergotropy:
\begin{equation}
    r_k = \mathcal{E}(t_{k+1}) - \mathcal{E}(t_k)
\end{equation}
where $\mathcal{E}(t_k)$ denotes the ergotropy of the battery at time $t_k$. This dense, immediate feedback directly rewards actions that increase ergotropy. The agent's objective is to maximize the undiscounted return $G = \sum_{k=0}^{K-1} r_k$, which telescopes to
\begin{equation}
    G = \mathcal{E}(t_{K}) - \mathcal{E}(t_{0})
\end{equation}
with the final time $t_K = \tau$. Thus maximizing $G$ is equivalent to maximizing the net increase in ergotropy. However, at early training stages the ergotropy-based reward is often zero, making it difficult to guide policy improvement. In practice, we therefore employ a weighted combination of energy gain and ergotropy gain as the reward. Following the weighting strategy in \cite{Erdman2024}, the scheme emphasizes energy gain when the ergotropy increment is typically zero, and progressively increases the weight on ergotropy gain as the policy improves:
\begin{equation}
\begin{aligned}
r_k ={}& w(n_{\mathrm{step}})(E(t_{k+1})-E(t_k)) \\
&+ (1-w(n_{\mathrm{step}}))(\mathcal{E}(t_{k+1})-\mathcal{E}(t_k)),
\end{aligned}
\label{eq:weighted_r}
\end{equation}
where the scheduled weight is chosen as
\begin{equation}
w(n_{\mathrm{step}})=\Bigl(1+e^{(n_{\mathrm{step}}-w_{\mathrm{mean}})/w_{\mathrm{width}}}\Bigr)^{-1}.
\label{eq:weight}
\end{equation}
Here $n_{\mathrm{step}}$ denotes the training step index, and $w_{\mathrm{mean}}$ and $w_{\mathrm{width}}$ are hyperparameters\cite{Erdman2024}. This schedule satisfies $w(n_{\mathrm{step}})\simeq 1$ at early training and $w(n_{\mathrm{step}})\to 0$ at late training, thereby shifting the reward from energy increments to ergotropy increments. While the default training uses the scheduled weight $w(n_{\mathrm{step}})$, we assess sensitivity to the reward mixing in Appendix \ref{app2} by comparing several fixed mixing values $w\in[0,1]$.

\subsection{Soft Actor–Critic algorithm}
We train the control policy with Soft Actor–Critic (SAC), an off-policy, entropy-regularized actor–critic method that alternates policy evaluation and policy improvement while maximizing a trade-off between return and action entropy \cite{Haarnoja2018}. Concretely, SAC optimizes
{\small
\begin{equation}
    \arg\max_{\pi} \mathbb{E}_{\pi}[\sum_{k=0}^{K-1}\gamma^{k}(r_k+\alpha\mathcal{H}[\pi(\cdot|s_k)]) | s_0=s_{\text{init}}]
\end{equation}
}
where $\gamma \in [0,1]$ is the discount factor and $\alpha \geq 0$ is the temperature controlling the exploration–exploitation. While our task is finite-horizon with an undiscounted terminal objective, we employ a discounted backup ($ \gamma<1 $) during training to reduce target variance and stabilize off-policy updates. 

\paragraph{Actor.}
The stochastic policy $\pi_\theta(a | s)$ outputs an unconstrained Gaussian sample
$
\tilde a(s) \sim \mathcal{N}\!\big(\mu_\theta(s),\,\sigma_\theta^2(s)\big),
$
which is pushed through a $\tanh$ map and then linearly rescaled to $[\lambda_{c,\min},\lambda_{c,\max}]$ (i.e., the squashed Gaussian policy in \cite{Erdman2024}):
{\small
\begin{equation}
  a(s) = \frac{\lambda_{c,\max}-\lambda_{c,\min}}{2}\,\tanh(\tilde a(s)) + \frac{\lambda_{c,\max}+\lambda_{c,\min}}{2}.
  \label{eq:squash-map}
\end{equation}
}
This yields differentiable, bounded actions compatible with the piecewise-constant control protocol of Sec.~\ref{sec:env}.

\paragraph{Critic and policy evaluation.}
Two state--action value networks $Q_{\phi_1}$ and $Q_{\phi_2}$ are learned to reduce over-estimation bias \cite{Haarnoja2018}. Given a replay buffer $\mathcal{B}$ of one-step transitions $(s,a,r,s')$, the critics minimize
{\small
\begin{equation}
  \mathcal{L}_Q(\phi_i) =
  \mathbb{E}_{(s,a,r,s')\sim \mathcal{B}}\!\left[
    \big(Q_{\phi_i}(s,a) - y\big)^2
  \right], i\in\{1,2\},
  \label{eq:critic-loss}
\end{equation}
}
with a clipped double-$Q$ target
{\small
\begin{equation}
  y = r + \gamma\,\mathbb{E}_{a'\sim\pi_{\theta}(\cdot|s')}\!\Big[
    \min_{j\in\{1,2\}} Q_{\phi^{\text{targ}}_j}(s',a')
    - \alpha\,\log \pi_{\theta}(a'| s')
  \Big],
  \label{eq:critic-target}
\end{equation}
}
and Polyak-averaged target networks
$
\phi^{\text{targ}}_j \leftarrow \rho\,\phi^{\text{targ}}_j + (1-\rho)\,\phi_j
$
with $\rho\in(0,1)$ \cite{Haarnoja2018,Erdman2024}.

\paragraph{Policy improvement.}
The actor is updated by minimizing \cite{Haarnoja2018,Erdman2024}:
{\small
\begin{equation}
  \mathcal{L}_\pi(\theta)
  =
  \mathbb{E}_{s\sim \mathcal{B},\;a\sim\pi_\theta(\cdot | s)}\!
  \Big[
    \alpha\,\log \pi_\theta(a | s)
    - \min_{j\in\{1,2\}} Q_{\phi_j}(s,a)
  \Big].
  \label{eq:actor-loss}
\end{equation}
}
The temperature $\alpha$ is adapted online to match a target entropy $\bar{\mathcal H}$ following \cite{Haarnoja2019, Erdman2024}:
{\small
\begin{equation}
  \mathcal{L}_{\text{temp}}(\alpha)
  =
  \mathbb{E}_{s\sim \mathcal{B},\;a\sim\pi_\theta(\cdot | s)}\!
  \big[
    \alpha\big(-\log \pi_\theta(a | s) - \bar{\mathcal H}\big)
  \big].
  \label{eq:temp-loss}
\end{equation}
}

\section{Simulation Results and Discussion}
\label{sec4}

\begin{figure}[htbp]
    \centering
    \subfigure[]{\includegraphics[width=0.9\columnwidth]{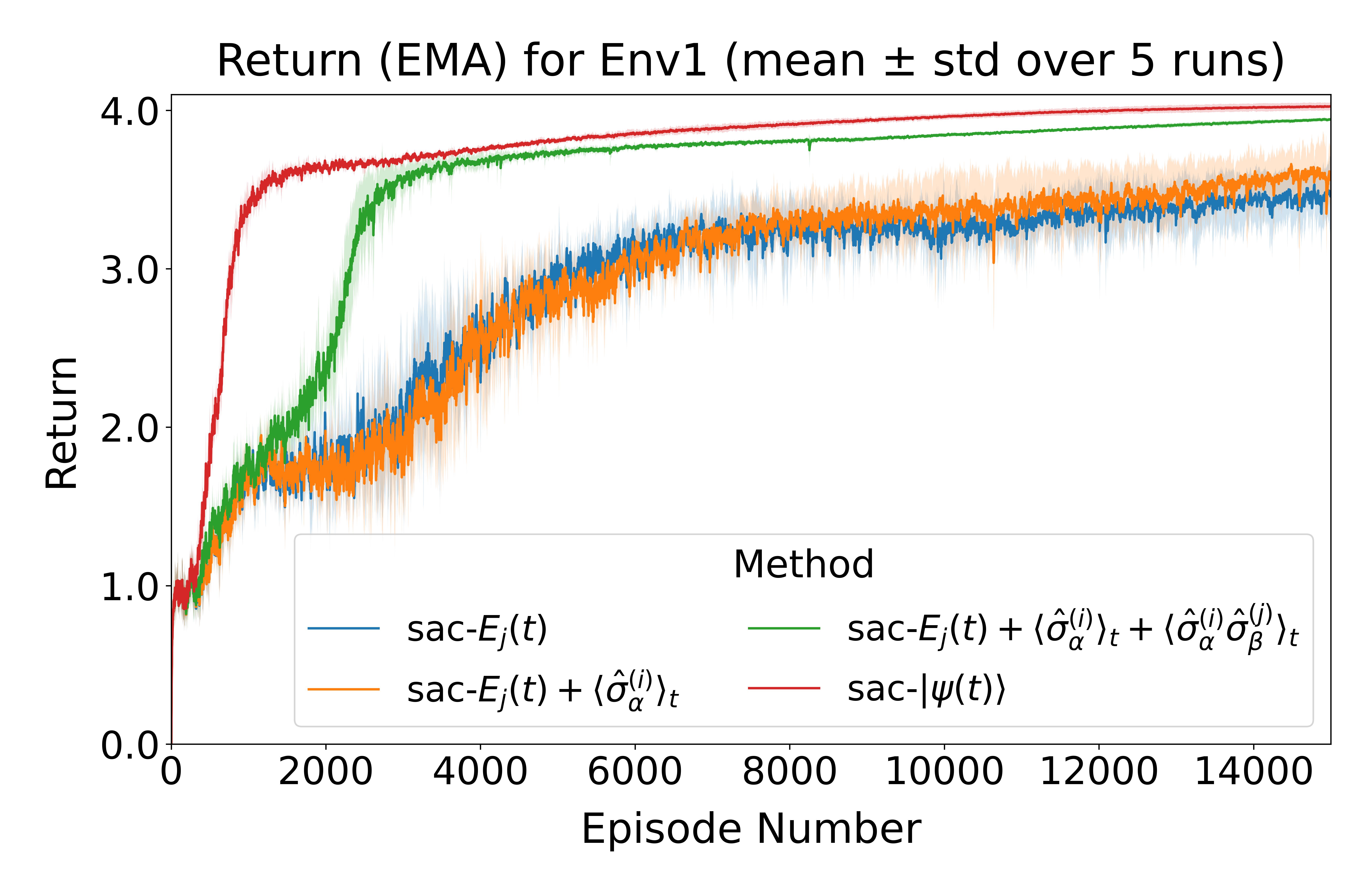}
    \label{figenv1_t}}
    \subfigure[]{\includegraphics[width=0.9\columnwidth]{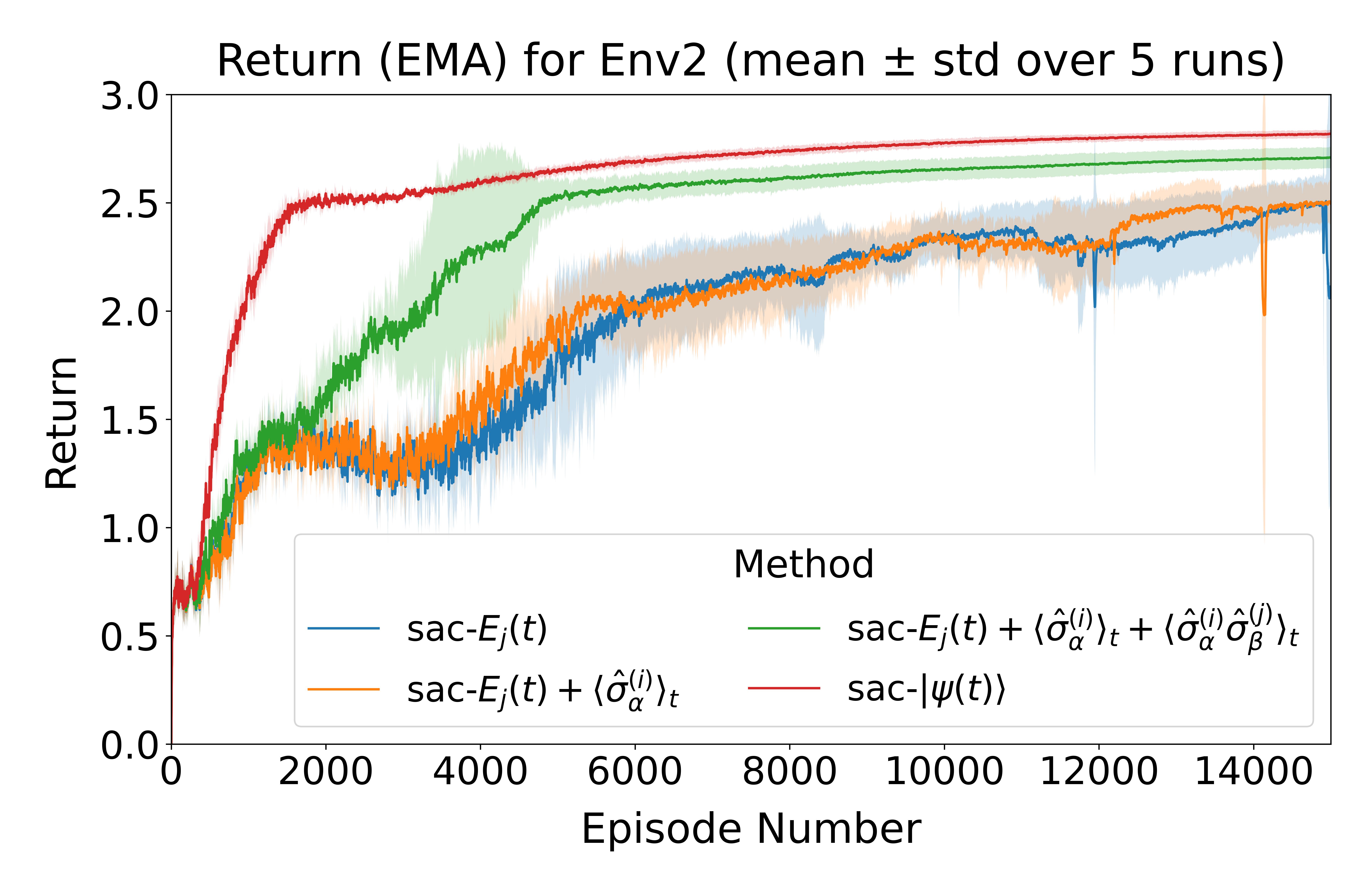}
    \label{figenv2_t}}
    \subfigure[]{\includegraphics[width=0.9\columnwidth]{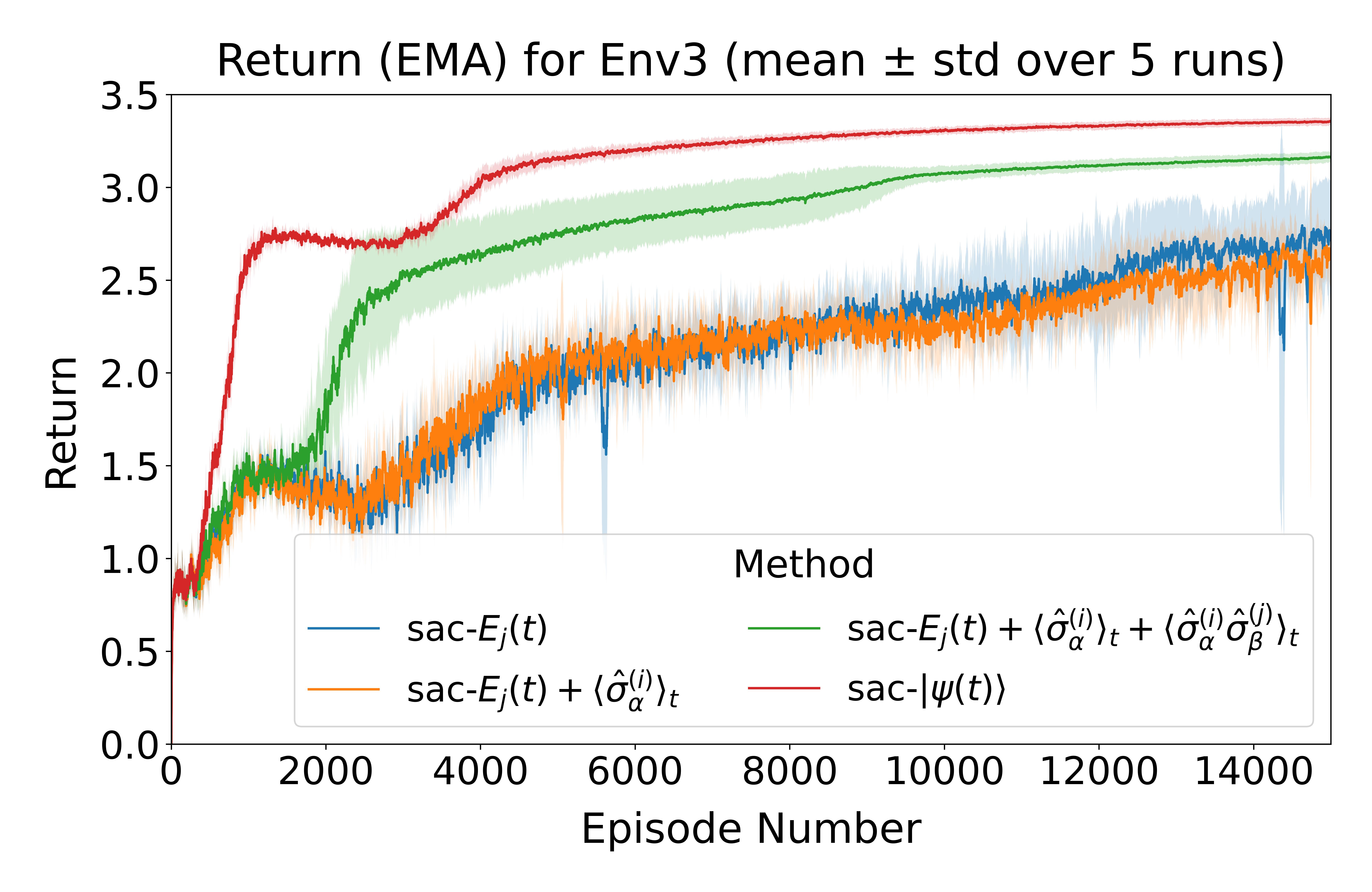}
    \label{figenv3_t}}
    \caption{Return (smoothed by an exponential moving average with $\gamma_G=0.9$) versus training episodes under four settings of available information, averaged over five random seeds, for three inhomogeneous quantum batteries: (a) Env1, (b) Env2, (c) Env3. The shaded regions represent the standard deviation over five seeds.}
    \label{fig:results_training}
\end{figure}
\begin{figure*}[t]
\centering
  \includegraphics[width=16cm]{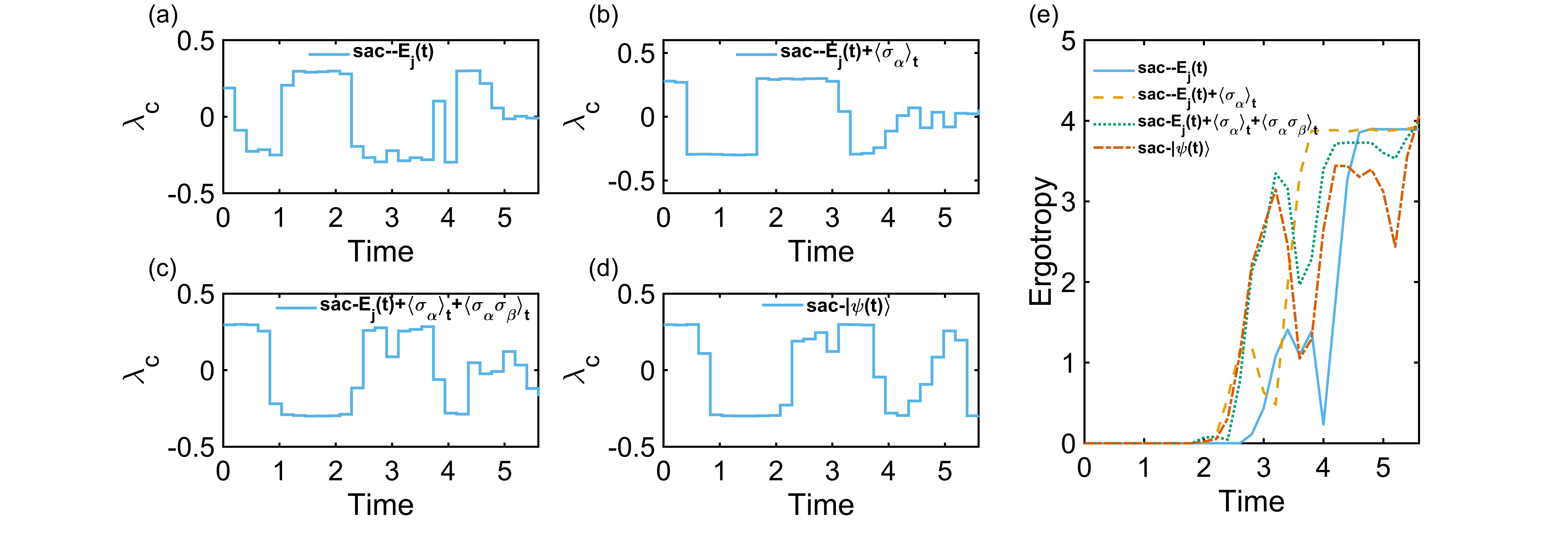}
\caption{Best charging protocols among five seeds for four our observable regimes [(a):$E_j(t)$, (b):$E_j(t) + \langle\hat{\sigma}_\alpha^{(i)}\rangle_t$, (c):$E_j(t) + \langle\hat{\sigma}_\alpha^{(i)}\rangle_t + \langle  \hat{\sigma}_\alpha ^{(i)}\hat{\sigma}_\beta^{(j)}\rangle_t$, (d):$|\psi(t)\rangle$] and their ergotropy trajectories (e) in Env1.}
\label{fig:results_protocol}
\end{figure*}

\subsection{Simulation parameters}
We simulate an inhomogeneous Dicke battery with $N=4$ TLSs. Inhomogeneity is introduced as static disorder in the TLS frequencies: the frequency of each TLS $\omega_j (j=1,2,3,4)$ is randomly sampled from a uniform distribution $U[0.5, 1.5]$ independently, referenced to a central frequency $\omega_0=1$. The total charging duration is set to $\tau=5.6$, discretized into $K=28$ equal steps of length $\Delta t = 0.2$. The agent controls the coupling strength, chosen continuously from the range $[-0.3, 0.3]$. Each experiment is repeated with five independent random seeds to assess variability.

Note that the disorder realization $\omega_j \sim U[0.5, 1.5]$ corresponds to relative static detunings up to $\delta_{\text{max}}=|\omega_j-\omega_0|/\omega_0 \leq 0.5$.To contextualize the physical relevance of the inhomogeneity parameters used in our simulations, we note that the inhomogeneous Dicke model is directly applicable to several experimental platforms where an ensemble of two-level systems is collectively coupled to a single bosonic mode, such as cavity/circuit QED systems and solid-state spin ensembles coupled to microwave resonators \cite{Baumann2010,MarquezPeraca2024,Remizov2018} . In such systems, typical inhomogeneous broadening—arising from factors like fabrication spreads in superconducting qubits or local field variations in spin ensembles—often ranges from $0.1\%$ to over $10\%$ of the transition frequency \cite{VanDamme2024,10.1063/1.4983350}. The broader disorder range examined here serves a dual purpose: it provides a conservative stress test for the reinforcement learning agent, verifying its ability to devise effective protocols even under significant detuning, and it accommodates potential additional detuning from environmental fluctuations or device aging, thereby demonstrating the robustness of the learned policies under realistically variable conditions.

\subsection{Experimental scenarios}
To assess the agent's ability to learn protocols under inhomogeneous conditions, we tested three disordered quantum battery systems with distinct qubit frequencies (independently sampled from $U[0.5, 1.5]$): (1) Env1: $[0.8745, 1.4507, 1.2320, 1.0987]$; (2) Env2: $[0.9170, 1.2203, 0.5001, 0.8023]$; (3) Env3: $[0.7220, 1.3707, 0.7067, 1.4186]$. Within each system we vary observability by exposing the agent to one of four observable regimes based on physical quantities:
\begin{enumerate}
    \item full state $|\psi(t)\rangle$;
    \item individual TLS expected energies $E_j(t)$;
    \item energies plus all possible first-order averages $\langle \hat{\sigma}_x^{(j)} \rangle_t$, $\langle \hat{\sigma}_y^{(j)} \rangle_t$ and $\langle \hat{\sigma}_z^{(j)} \rangle_t$ with  $j\in\{1,2,\cdots,N\}$;
    \item energies plus first-order averages and all possible second-order correlations $\langle  \hat{\sigma}_\alpha ^{(i)} \hat{\sigma}_\beta^{(j)}\rangle_t$ with $\alpha,\beta\in\{x,y,z\}$ and $i,j\in\{1,2,\cdots,N\}$.
\end{enumerate}

During training we log the ergotropy achieved by all charging protocols explored by the agent and, for each setting, report the best value attained (i.e., the maximum over explored protocols). 

\subsection{Performance analysis}

\begin{table*}[htbp]
\centering
\caption{Mean and 95\% confidence interval (CI, using Student-$t$ interval)} of the best terminal ergotropy attained during training across five seeds for each setting of available information in Env1–Env3.
\label{tab:results}
\begin{tabular*}{\textwidth}{@{\extracolsep{\fill}}lccc}
\toprule
\multirow{2}{*}{\textbf{Available Information}}  
  & \multicolumn{3}{c}{\textbf{Ergotropy (mean $\pm$ CI half-width)}} \\
  \cmidrule(lr){2-4}
  & \textbf{Env1} & \textbf{Env2} & \textbf{Env3}  \\
\midrule
$|\psi(t)\rangle$
& 4.0445 $\pm$ 0.0323 & 2.8357 $\pm$ 0.0235 & 3.3777 $\pm$ 0.0235
\\ 
$E_j(t)$ 
& 3.8496 $\pm$ 0.0459 & 2.6757 $\pm$ 0.0902 & 3.0938 $\pm$ 0.1648
\\
$E_j(t) + \langle\hat{\sigma}_\alpha^{(i)}\rangle_t$
& 3.8984 $\pm$ 0.0326 & 2.6548 $\pm$ 0.0516 & 3.0585 $\pm$ 0.0519
\\
$E_j(t) + \langle\hat{\sigma}_\alpha^{(i)}\rangle_t + \langle  \hat{\sigma}_\alpha ^{(i)} \hat{\sigma}_\beta^{(j)}\rangle_t$
& 3.9683 $\pm$ 0.0139 & 2.7260 $\pm$ 0.0575 & 3.1966 $\pm$ 0.0397
\\ \bottomrule
\end{tabular*}
\end{table*}

The numerical results are summarized in Table~\ref{tab:results} and the training processes are visualized in Fig.~\ref{fig:results_training}, where the return is smoothed using an exponential moving average (EMA) with coefficient $\gamma_G$, defined as $\bar{G}_t = \gamma_G \bar{G}_{t-1} + (1-\gamma_G) G_t$, where $G_t$ is the return at episode $t$, $\bar{G}_t$ is the EMA-smoothed return, and $\gamma_G \in [0,1)$ controls the smoothing strength. Fig.~\ref{fig:results_protocol} further displays, for Env1, the best charging protocol identified among the five seeds and the corresponding ergotropy trajectory. Additional metrics across seeds, including distribution of final ergotropies and convergence rates, are provided in Appendix~\ref{app1}. From these results, we draw several key findings.

Under full observability, the agent attains the highest possible ergotropy, learning near-optimal charging protocols. It also exhibits the smallest variability across seeds, as evidenced by the narrowest shaded region in Fig.~\ref{fig:results_training}. This indicates that access to the complete state enables the agent to make consistent, high-quality control decisions.

Under more realistic, partially observable conditions, a clear performance hierarchy emerges with increasing information content of the observables: $[E_j(t)] \approx [E_j(t) + \langle\hat{\sigma}_\alpha^{(i)}\rangle_t] < [E_j(t) + \langle\hat{\sigma}_\alpha^{(i)}\rangle_t + \langle  \hat{\sigma}_\alpha ^{(i)} \hat{\sigma}_\beta^{(j)}\rangle_t]$. Observing only individual TLS energies is insufficient to learn an effective protocol, and adding first-order averages offers negligible—if any—improvement. However, including correlation features recovers a substantial fraction of the performance achieved under full observability: across Envs 1–3, the average best ergotropy (over 5 seeds) reaches 98.11\%, 96.13\%, and 94.64\% of the fully observed baseline, respectively.

The performance gap across observability regimes is consistent with aliasing in POMDPs: restricted observables can map distinct quantum states to similar observations, resulting in suboptimal actions and increased variance across training runs. Incorporating second-order correlations enriches the observation with information about inter-spin correlations, which helps mitigate aliasing and improves both terminal performance and training stability.

From Fig.~\ref{fig:results_protocol}(c) and~\ref{fig:results_protocol}(d), enriching the observables with both first-order averages and second-order correlations yields a charging protocol that closely matches the one learned under full-state access. Moreover, Fig.~\ref{fig:results_protocol}(e) shows that an effective protocol does not maximize ergotropy at every instant; the agent selects the coupling strength to maximize terminal ergotropy rather than immediate gains.

\paragraph{Robustness across disorder realizations.}
To further assess robustness of the learned RL agent beyond the training realization, we train the agent on Env1 and periodically evaluate it on Env1--Env3 without retraining. We select, for each seed, the checkpoint that maximizes the mean terminal ergotropy across Env1--Env3, and report the corresponding terminal ergotropy achieved on each environment. The resulting robustness evaluation averaged over five random seeds is summarized in Table~\ref{tab:robust}.

\begin{table*}[htbp]
\centering
\caption{Evaluation result of the learned RL agent (trained on Env1) across disorder realizations (Env1--Env3). We select, for each seed, the checkpoint that maximizes the mean terminal ergotropy across Env1--Env3, and report the corresponding terminal ergotropy achieved on each environment (Mean and 95\% confidence interval).}
\label{tab:robust}
\begin{tabular*}{\textwidth}{@{\extracolsep{\fill}}lccc}
\toprule
\multirow{2}{*}{\textbf{Available Information}}  
  & \multicolumn{3}{c}{\textbf{Ergotropy(mean $\pm$ CI half-width)}} \\
  \cmidrule(lr){2-4}
  & \textbf{Env1} & \textbf{Env2} & \textbf{Env3}  \\
\midrule
$|\psi(t)\rangle$
& 4.0024 $\pm$ 0.0246 & 2.6178 $\pm$ 0.1222 & 3.1248 $\pm$ 0.0905
\\ 
$E_j(t)$ 
& 3.4844 $\pm$ 0.3115 & 0.9452 $\pm$ 0.6414 & 1.9187 $\pm$ 0.7559
\\
$E_j(t) + \langle\hat{\sigma}_\alpha^{(i)}\rangle_t$
& 3.6658 $\pm$ 0.1750 & 0.9623 $\pm$ 0.6043 & 2.2375 $\pm$ 0.7476
\\
$E_j(t) + \langle\hat{\sigma}_\alpha^{(i)}\rangle_t + \langle  \hat{\sigma}_\alpha ^{(i)} \hat{\sigma}_\beta^{(j)}\rangle_t$
& 3.8776 $\pm$ 0.0537 & 1.7048 $\pm$ 0.8825 & 2.7069 $\pm$ 0.2357
\\ \bottomrule
\end{tabular*}
\end{table*}

Across all three disorder realizations, we observe a consistent ordering of robustness:
full-state access yields the strongest cross-realization performance, while energy-only observations generalize the least. Augmenting the observation with first-order moments provides only a limited improvement, whereas including second-order correlations produces a clear robustness gain, consistent with the key role of correlation information under inhomogeneity. Nevertheless, a noticeable robustness gap to the full-state reference remains after adding correlations, especially in Env2, indicating that two-body information, while highly informative, is not sufficient to reach full-state-level robustness across realizations in the present setting.







\subsection{Discussion}

Our results confirm that reinforcement learning is an effective approach for learning high-quality charging protocols in inhomogeneous quantum batteries. When full state information (i.e., $|\psi(t)\rangle$) of the system is available, the agent consistently discovers ideal solutions and exhibits stable learning dynamics.

The more consequential finding concerns the practical scenario of partial observability. We show that the choice of observables is critical, with second-order correlations playing an essential role in attaining high charging performance. In particular, correlation measurements largely compensate for the information gap between single-TLS energy readouts and full-state access, thereby enabling effective optimization.

In addition, an intriguing observation from our results is that first-order averages contribute little to estimation performance, whereas second-order correlations yield a substantial gain. This can be physically explained by a conserved discrete parity symmetry in our model. The symmetry is generated by the operator $\hat{P}=(\otimes_{j=1}^Ni\hat{\sigma}_z^{(j)})\otimes e^{i\pi\hat{a}^\dagger \hat{a}}$. Under this parity, single-spin operators $\hat{\sigma}_x^{(j)}$ and $\hat{\sigma}_y^{(j)}$ are odd ($\hat{P} \hat{O}\hat{P}^\dagger=-\hat{O}$), while $\hat{\sigma}_z^{(j)}$ is even. Since the initial state is a parity eigenstate and the Hamiltonian commutes with $\hat{P}$, the expectation values of all odd single-spin operators remain exactly zero throughout the evolution. Consequently, their first-order averages provide no information beyond the energy, which depends solely on the even observable $\langle \hat{\sigma}_z^{(j)} \rangle$. In contrast, second-order correlators such as $\langle \hat{\sigma}_x^{(i)}\hat{\sigma}_x^{(j)} \rangle$ are products of two odd operators, resulting in an even composite observable that can take nonzero values. These correlators thus encode genuine multi-body quantum correlations that are fundamentally absent from the first-order data, accounting for the significant performance improvement when they are included.

The finding that second-order correlations are nearly sufficient for ergotropy optimization, while surprising from a quantum-state-tomography standpoint, has a physical basis. Ergotropy, as a macroscopic observable, does not require full state reconstruction. Its optimization depends predominantly on global energy distributions and collective order, which are typically governed by low-order correlations in physical systems. Consequently, the agent's ability to achieve near-optimal charging using only these correlations is thus inherent to the physical task itself, not an artifact of the specific RL policy.

Finally, the ergotropy trajectories indicate that effective charging is inherently nonmyopic. The learned agent does not chase instantaneous ergotropy gains; instead, it schedules coupling strengths to improve the terminal outcome, even at the cost of temporary plateaus or counterintuitive moves.

In the correlation observability regime, the agent is provided with the full set of two-body Pauli--Pauli expectation values $\{\langle \sigma_\alpha^{(i)}\sigma_\beta^{(j)}\rangle_t\}$, in addition to the single-body information.  While a naive counting suggests up to $9N^2$ distinct correlators, physical constraints, including parity and exchange symmetry drastically reduce this number to just $5N(N-1)/2$ independent measurements, a reduction of approximately $72\%$ from the naive scaling. For larger systems, this quadratic overhead can be further mitigated. Techniques from compressed sensing enable the reconstruction of all correlations from a number of random measurements scaling only linearly or sub-quadratically with $N$ \cite{PhysRevLett.105.150401}. Alternatively, a recurrent neural network policy could be trained on a time-series of easily measurable local observables, implicitly estimating the correlations within a partially observable framework \cite{Kaelbling1998}. For the moderate system size studied here, and given the capability for parallel readout in platforms like superconducting qubits or trapped ions, the measurement set is experimentally manageable, ensuring the practical relevance of our approach.

\section{Conclusions}
\label{sec5}

In this work, we formulated fast charging of an inhomogeneous Dicke quantum battery as a discrete-time reinforcement learning problem and trained piecewise-constant control policies with Soft Actor–Critic algorithm. The agent optimizes terminal ergotropy using a dense, shaped reward that blends energy and ergotropy gains, and acts within a bounded, continuous action space for the charger–battery coupling strength. To reflect realistic readout constraints, we examined four observability regimes—from full state access to experimentally accessible summaries built from single-TLS energies, first-order averages, and second-order correlation features.

Across three disordered four-TLS quantum batteries, SAC consistently discovers high-quality charging protocols. Under full observability, the agent attains near-optimal terminal ergotropy with low seed-to-seed variability. Under partial observability, a clear hierarchy emerges: access to only single-TLS energies ($E_j(t)$) or energies plus first-order averages ($E_j(t) + \langle\hat{\sigma}_\alpha^{(i)}\rangle_t$) lags behind the fully observed baseline, whereas augmenting with second-order correlations ($E_j(t) + \langle\hat{\sigma}_\alpha^{(i)}\rangle_t + \langle  \hat{\sigma}_\alpha ^{(i)}\hat{\sigma}_\beta^{(j)}\rangle_t$) recovers most of the gap, reaching 94\%–98\% of the full-state performance depending on the battery parameters. Analysis of the learned schedules further shows that effective charging is nonmyopic: the agent modulates the coupling to maximize terminal ergotropy rather than pursuing instantaneous gains, accepting temporary plateaus or declines to achieve superior final outcomes.

These results establish RL as a practical tool for optimizing fast-charging protocols in inhomogeneous quantum batteries under information constraints and point to several promising extensions: (i) Adopting more principled POMDP approaches — for example, equipping the agent with recurrent architectures that use observation histories — could help the policy form an internal belief over latent quantum states, thereby handling partial observability induced by compressed, experimentally accessible readouts. (ii) Seeking robust fast-charging protocols under per-episode fluctuations of TLS frequencies within specified ranges can improve reliability when device parameters drift between charges. Together, these extensions would enhance practicality by addressing both information limitations and parameter variability in experimental quantum batteries.

\appendix

\section{Additional metrics across seeds}
\label{app1}

\renewcommand{\thefigure}{S\arabic{figure}}
\renewcommand{\thetable}{S\arabic{table}}
\setcounter{figure}{0}
\setcounter{table}{0}

\begin{figure}[htbp]
    \centering
    \subfigure[]{\includegraphics[width=0.9\columnwidth]{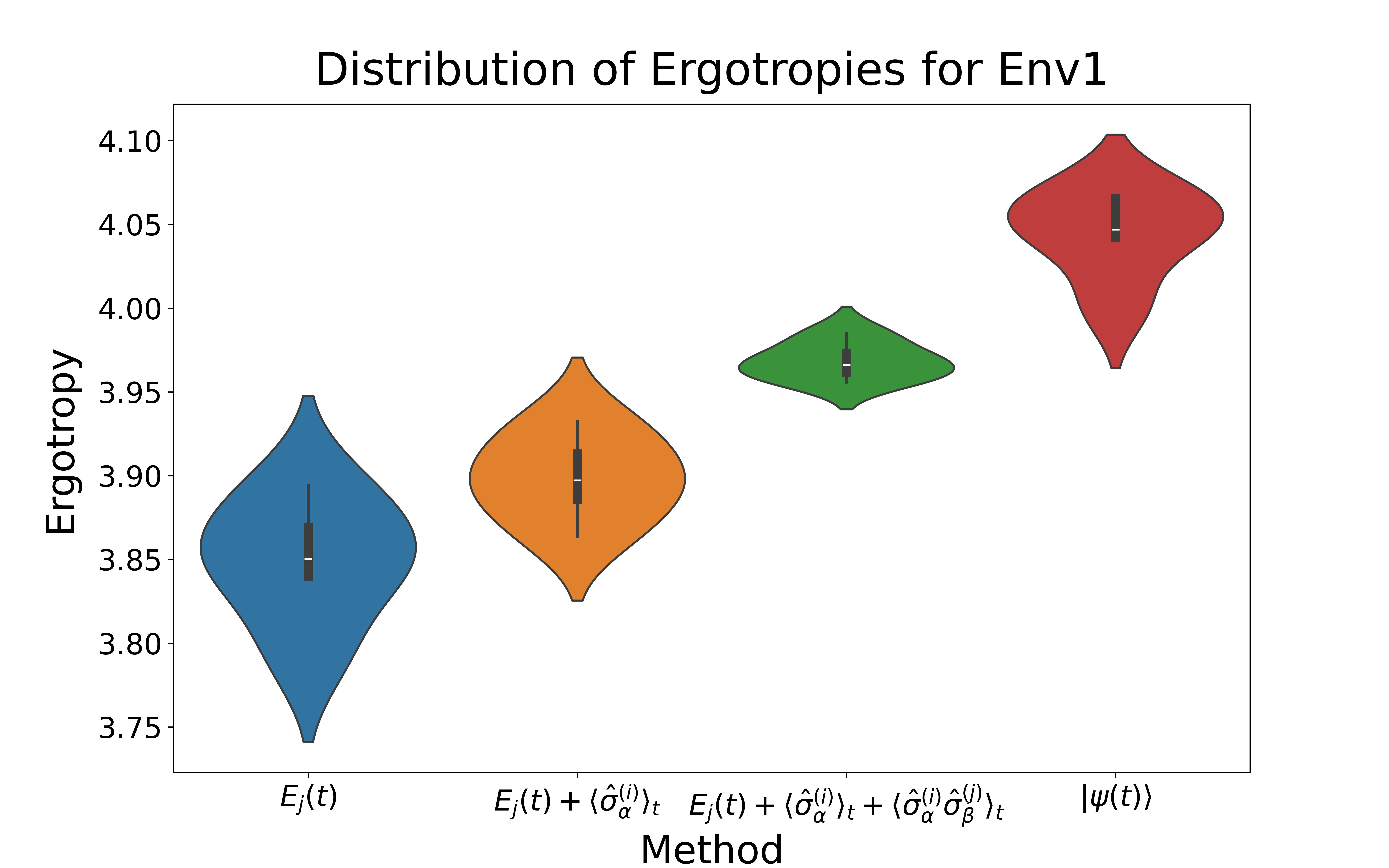}
    \label{figenv1_dis}}
    \subfigure[]{\includegraphics[width=0.9\columnwidth]{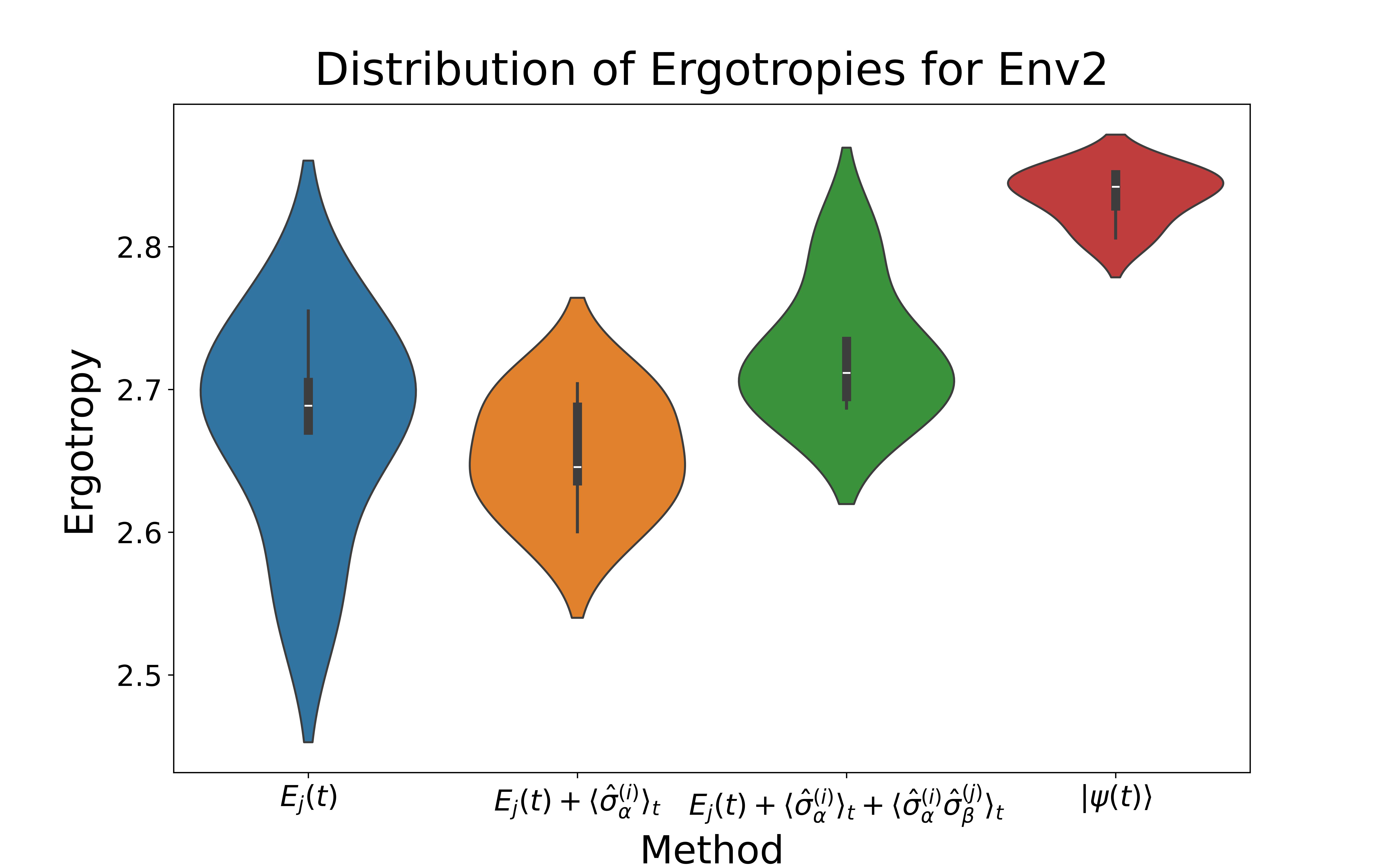}
    \label{figenv2_dis}}
    \subfigure[]{\includegraphics[width=0.9\columnwidth]{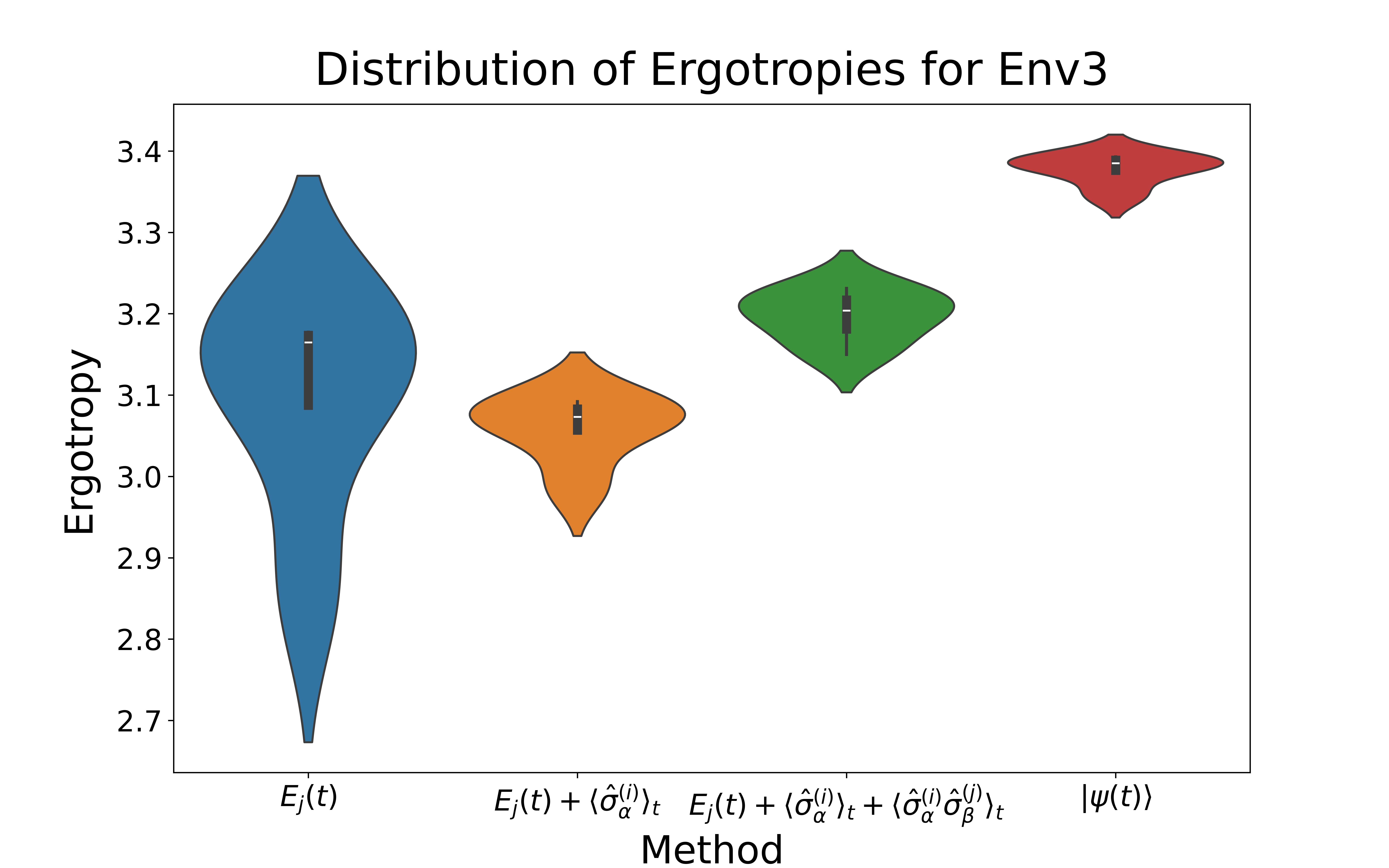}
    \label{figenv3_dis}}
    \caption{The distribution of the best terminal ergotropies attained during training under four settings of available information over five random seeds, for three inhomogeneous quantum batteries: (a) Env1, (b) Env2, (c) Env3.}
    \label{fig:results_dis}
\end{figure}

\begin{table*}[t]
\caption{Convergence rate across five seeds.
$T_{0.9}$ denotes the number of training episodes required for the terminal ergotropy attained during training to first reach $90\%$ of its final best value.}
\label{tab:conv}
\begin{tabular*}{\textwidth}{@{\extracolsep{\fill}}lccc}
\toprule
\multirow{2}{*}{\textbf{Available Information}}  
  & \multicolumn{3}{c}{\textbf{$T_{0.9}$}(mean $\pm$ CI half-width)} \\
  \cmidrule(lr){2-4}
  & \textbf{Env1} & \textbf{Env2} & \textbf{Env3}  \\
\midrule
$|\psi(t)\rangle$
& 948.8 $\pm$ 166.6 & 1474.8 $\pm$ 374.0 & 3209.2 $\pm$ 190.8
\\ 
$E_j(t)$ 
& 1744.0 $\pm$ 390.8 & 1631.8 $\pm$ 418.5 & 3436.2 $\pm$ 689.0
\\
$E_j(t) + \langle\hat{\sigma}_\alpha^{(i)}\rangle_t$
& 2039.8 $\pm$ 499.0 & 1700.6 $\pm$ 276.1 & 3653.4 $\pm$ 1185.5
\\
$E_j(t) + \langle\hat{\sigma}_\alpha^{(i)}\rangle_t + \langle  \hat{\sigma}_\alpha ^{(i)} \hat{\sigma}_\beta^{(j)}\rangle_t$
& 1921.4 $\pm$ 326.0 & 2097.0 $\pm$ 610.6 & 4350.4 $\pm$ 2530.3
\\ \bottomrule
\end{tabular*}
\end{table*}

Fig.~\ref{fig:results_dis} summarizes the distribution of the best terminal ergotropy attained during training under four observability settings for Env1--Env3. Across all disorder realizations, access to the full state consistently yields the highest terminal ergotropy and, at the same time, typically exhibits the most concentrated distributions, indicating both superior achievable performance and strong robustness to random initialization. In contrast, restricted observations lead to broader spreads and lower typical values. Notably, augmenting the observation with second-order correlations produces a clear and systematic improvement over energy-only and first-order information: the distributions shift toward higher ergotropy, in most cases, become more concentrated, reflecting a more reliable optimization outcome across seeds. Nevertheless, a non-negligible gap to the full-state reference remains, consistent with the partial-observability limitations discussed in the main text.

Table~\ref{tab:conv} quantifies the convergence rates. We define $T_{0.9}$ as the number of training episodes required for the best-so-far terminal ergotropy to first reach $90\%$ of its final best value. As expected, full-state access generally leads to the fastest optimization (smallest $T_{0.9}$), reflecting more informative feedback for learning. By comparison, adding second-order correlations does not necessarily accelerate convergence relative to lower-order observations and can even lead to slower convergence in some environments; however, it yields substantially better terminal ergotropy as shown
in Fig.~\ref{fig:results_dis} and Table~\ref{tab:results}. In our context, this trade-off is favorable: the primary objective is to maximize the extractable work at the end of charging, and the correlation-augmented setting achieves markedly improved final performance even when the convergence rate is not improved.

\section{Sensitivity to the reward-mixing weight}
\label{app2}

\begin{figure}[t]
\includegraphics[width=\linewidth]{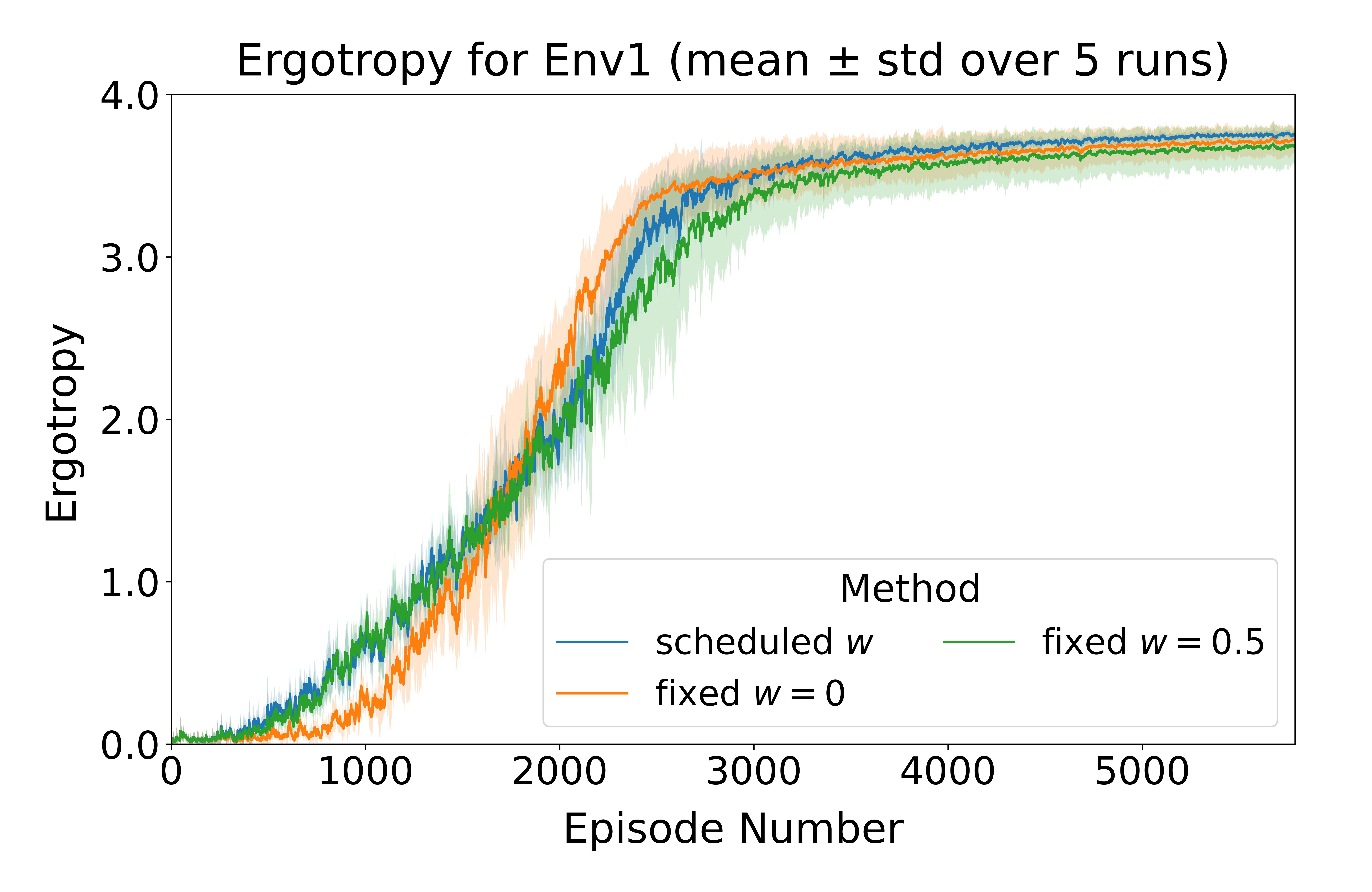}
\caption{Sensitivity to the reward-mixing weight. Comparison of scheduled $w$ with fixed $w=0.5$ (mixed reward) and $w=0$ (ergotropy-only reward) in $r_k=w\,\Delta E_k+(1-w)\Delta\mathcal{E}_k$.
All agents are trained on Env1 with the correlation-augmented observation $E_j(t)+\langle \hat{\sigma}^{(i)}_\alpha\rangle_t+\langle \hat{\sigma}^{(i)}_\alpha \hat{\sigma}^{(j)}_\beta\rangle_t$. Curves show the ergotropy during training (EMA smoothing with $\gamma=0.9$); shaded regions indicate the standard deviation over five seeds.}

\label{fig:reward_weight}
\end{figure}

The main text employs a shaped reward that mixes the energy and ergotropy increments $
r_k = w\,\Delta E_k + (1-w)\,\Delta \mathcal{E}_k$, where $w$ is either fixed or scheduled during training according to Eq. \eqref{eq:weighted_r} and \eqref{eq:weight}, motivated by the observation that ergotropy increments can be sparse at early training stages. To assess how sensitive learning is to the reward-weighting choice, we perform an ablation study on the reward mixing under the same environment and observation setting used for our main partial-observability results. Specifically, all agents in this appendix are trained on Env1 using the correlation-augmented observation $E_j(t)+\langle \hat{\sigma}^{(i)}_\alpha\rangle_t+\langle \hat{\sigma}^{(i)}_\alpha \hat{\sigma}^{(j)}_\beta\rangle_t$, and differ only in the reward-weighting scheme. The corresponding learning curves are shown in Fig.~\ref{fig:reward_weight}.

The comparison highlights the distinct roles of the energy and ergotropy terms at different stages of learning. At early training, increments in ergotropy are frequently zero (or extremely small), making $\Delta\mathcal{E}_k$ a sparse learning signal. Assigning a nonzero weight to the energy increment therefore provides denser feedback and accelerates the initial growth of the achieved ergotropy. As exploration proceeds and nontrivial ergotropy gains become more common, increasing the relative weight on $\Delta\mathcal{E}_k$ guides the agent toward more goal-directed exploration and leads to faster improvement in ergotropy. Consistent with this interpretation, the scheduled-weight strategy---which interpolates between these regimes---achieves the best overall trade-off and attains the highest ergotropy at late training compared with either fixed weighting.

\section*{Acknowledgements}
This work is supported by the Tsinghua-Toyota Joint Research Fund.

\bibliographystyle{unsrt}
\bibliography{QB}

\end{document}